\documentclass[journal]{IEEEtran}
\usepackage{epsfig}
\usepackage{epstopdf} 
\usepackage{cite}
\usepackage[cmex10]{amsmath}
\usepackage{algorithmic}
\usepackage{array}
\usepackage{mdwmath}
\usepackage{mdwtab}
\usepackage{subfigure}
\usepackage{color}
\usepackage{amsmath}

\usepackage{boxedminipage}

\begin{document}

\title{Femtocaching and Device-to-Device Collaboration: A New Architecture for Wireless Video Distribution}

\author{Negin Golrezaei,~\IEEEmembership{Student Member,~IEEE,}
        Andreas F.~Molisch,~\IEEEmembership{Fellow,~IEEE,}
        Alexandros G.~Dimakis,~\IEEEmembership{Member,~IEEE,}
        and~Giuseppe~Caire,~\IEEEmembership{Fellow,~IEEE}
        }

\markboth{}%
{}

\maketitle

\begin{abstract}

We present a new architecture to handle the ongoing
explosive increase in the demand for video content in wireless networks. It is based on distributed caching of the content in femto-basestations with small or non-existing backhaul capacity but with considerable storage space, called \emph{helper nodes}. We also consider using the mobile terminals themselves as caching helpers, which can distribute video through device-to-device communications.  This approach allows an
improvement in the video throughput without deployment of any
additional infrastructure. The new architecture can improve video throughput by one to two orders-of-magnitude.  

\end{abstract}

\IEEEpeerreviewmaketitle

\section{Introduction}

\IEEEPARstart{V}{ideo} transmission is currently the main driver for the increase of both wired and wireless data traffic.  In wired networks, movie streaming currently accounts for $50\%$ of all internet traffic during evening hours and further growth is expected. A similar trend is observed for wireless networks. Wireless data traffic is expected to increase by a factor of $40$ over the next five years, from currently $93$ Petabytes to $3600$ Petabytes per month~\cite{cisco66}. This explosive demand is fueled mainly~\cite{cisco66} by video traffic, which is expected to increase by a factor of $65$ times, and become the by far dominant source of data traffic. 
This trend is accelerated by the proliferation of mobile devices that allow comfortable viewing experiences (tablet computers, large-screen phones), the higher content quality, and the proliferation of social networks. Concurrently, 
the emergence of on-demand video streaming services involves repeated wireless transmission of videos
that are viewed multiple times by different users in a completely asynchronous way.
Since the relative delay at which two users may stream the same video is generally
much larger than the duration of the video itself, a conventional network architecture treats
each streaming session as independent data, and it is therefore incapable of exploiting
either the redundancy in the demands (the same file requested over and over) or the intrinsic multicasting capability of the wireless 
medium (unlike live streaming, for which the multicasting capability has been exploited in several 
system proposals and system implementations such as MediaFLO). 

This increasing demand offers new business opportunities but also poses an enormous challenge from a technical and economical perspective. 
Anecdotal user experience, as well as statistics from many countries, show that current wireless networks are bursting at the seams.
As demand is growing, 
operators have to constantly reduce the cost per transferred bit without relying on increasing wireless spectrum. 
Absent such developments, traffic--decreasing measures (caps on data per user, throttling the data rate of high--usage customers) 
would have to be taken, stymying the positive developments. 

Traditional methods for increasing data throughput in wireless networks have relied on the following three approaches: (i) increase of spectrum usage, (ii) increase of the per--link spectral efficiency, and (iii) increase of 
spatial reuse. Increasing the amount of spectrum is limited by the fact that spectrum is a finite resource, and allocating new bands to cellular services is a long and expensive process. Increasing the spectral efficiency per link is also approaching its limits: fourth--generation cellular systems such as LTE have a near--optimal physical layer, using OFDM together with capacity--approaching  codes and multiple antenna elements. While further improvements will be possible through  novel interference management techniques (cooperative multipoint, interference alignment, etc.), such solutions are expensive and/or require redesign of the physical layer, which in turn needs a new cellular standard.
This leaves as the main measure the decrease of distance between transmitter and receiver, and thus the increase of the area spectral efficiency. Indeed, this development has been widely recognized, and forms the motivation for the current high interest in femtocells and their integration into cellular networks to form large--scale heterogeneous networks. This approach --which can be combined with increased spectrum and more efficient links --is highly scalable, and can thus provide the order--of-magnitude increase in capacity that is required for future video networks. The Achilles heel lies in the fact that every Base Station (BS) --including femto-stations --needs a high-speed backhaul, whose quality must be better than the aggregate data rate of all its served users.
Optical fiber can provide the capacity, but bringing a fiber-connection to the location of each small, possibly user-deployed, base station, might be labor-intensive and thus expensive. On the other hand, DSL or microwave backhaul may not meet  the rate requirements and may become the system bottleneck
(e.g., the current average DSL download speed in USA is $\sim$ 5 Mb/s, while a WLAN can achieve 30 Mb/s and more). 

In this paper, we are describing a radically new approach that is based on the following two key observations: (i) a large amount of video traffic is caused by a few, popular, files and (ii) disk storage is a quantity that increases faster than any other component in communications/processing systems. As a matter of fact, typical hard disks have increased storage space from $30$ MByte (in $1991$) to $3$ TByte (in $2011$) -- an increase of five orders of magnitude. The capabilities of storage devices have been further enhanced by improved coding for storage, in particular excellent distributed storage codes.  The essential idea of our approach thus is to trade off backhaul capacity with caching of video files at local base stations or helper nodes from where they can be transmitted very efficiently. In other words, we envision a proliferation of low-complexity base stations with weak backhaul links, which we call \emph{helper nodes} \cite{zhang2010adaptive}. Those nodes obtain the most popular video files by downloading through their weak backhaul links (or even completely without backhaul, through methods discussed later on). Whenever a user close to them needs a video file that they have stored, they transmit to it over a wireless link that is short-distance and thus has extremely high area spectral efficiency. Even though the helper nodes clearly cannot store all possible video files, the approach is very efficient, because a small subset of video files accounts for the majority of video traffic. For those files that are not cached in the helpers at all (because they are not popular enough), download via the traditional macrocellular base stations is used. Depending on the popularity distributions of the video files and the cache sizes, a large percentage of video traffic can be offloaded to the helper system.

One key question for such a system is the wireless distributed caching problem, \textit{i.e.}, which files should be cached by which helpers. If every mobile device has only access to a exactly one helper, then clearly each helper should cache the same files, namely the most popular ones. However, for the case that each mobile device can access multiple caches, the assignment of files to helpers becomes nontrivial. We will outline below a formulation of the optimal caching problem. While an exact solution is NP-hard, approximate solutions whose performance is provably within $50 - 63\%$ of the optimum, can be found. We will also discuss how the popularity distribution of video files can be learned and/or predicted, and the content of the caches changed according to the variations in the popularity distribution.

Our concept is pushed even further by introducing the notion of mobile devices themselves as helper nodes. Recent years have seen an enormous proliferation of smartphones and tablets that have anywhere between $10$ and $64$ GByte of storage (not to mention the $500$ GByte on typical laptop hard disks). By enabling device-to-device communications, the ensemble of mobile devices can become a distributed cache that allows a much more efficient downloading. The advantage of using mobile devices instead of fixed helper nodes lie in the small deployment costs and automatic upscaling of the capacity as the density of such devices increases. The drawback lies in the necessity to motivate users to participate in the cache, and the more random nature of the available throughput. We envision that as a video-infrastructure is built out, the system progresses from opportunistic caching in the User Terminals (UTs) (requires minimal infrastructure changes), to the deployment of distributed helper systems, to systems that additionally employ device-to- device communications. 

For mobile devices as caches, some of the fundamental questions are (i) over what distance should device-to-device communication be enabled? (ii) does a centralized control of the communications (\textit{i.e.}, a macro base station determines which devices communicate with each other though the payload data exchange happens without the BS) have significant advantages? (iii) what files should be stored by the mobile devices? and (iv) how can they most efficiently acquire those files? 

In the subsequent sections we will describe more technical details of our approach, and provide simulation results of the performance enhancements. We will show that an orders-of-magnitude increase of video throughput is possible. While many of the results can be derived in closed form, we refer to our recent research papers \cite{website}, for details. 

\section{Distributed storing of video and popularity distributions}

The capacity of commercial hard disks has increased dramatically over the past years. Currently (late $2011$), storage capacity of $2-3$ TBytes is available at retail prices around $100$ US \$.  Thus, it is eminently feasible to build several TByte of storage into a femto-cell base station at low cost. At the same time, modern smartphones and tablets have significant storage capacity often reaching several gigabytes. Recent breakthroughs in dense NAND flash are making 128GB smartphone memory chips available~\cite{flash128}.
Significant advances have been made in the area of storage coding. While earlier work had mostly concentrated on protecting data stored on a single storage node, recent work has resulted in efficient distributed storage codes \cite{dimakis2010network}. In other words, redundant bits are distributed over multiple storage nodes, in such a way that files can be reconstructed even in the case of multiple failures. 
Erasure coding techniques such as distributed storage codes and fountain codes allow the recovery of an original video file as long as the number of encoded bits available at the receiver is larger than or equal to the number of original source bits. The benefit of coding can be seen as an analogy of converting source bits into water. It does not matter  which specific coded bits are available as long as a receiver collects enough information by filling a bucket with water from a fountain: the identity of which water droplets are collected does not play a role.  

Distortion-aware Storage codes can be also designed \cite {dimakis2007unequal} building on scalable video 
source codes. The main idea is that videos can be encoded at different quality levels, which in turn require different amounts of bits. In other words, a particular video sequence can be compressed significantly, and then be played back with reduced clarity and sharpness, if only a small bandwidth is available. Alternatively, if more bandwidth can be used, additional information that can refine the picture quality, can be employed. This principle is known as multi-level coding (also known as  successive refinement, multi-resolution, and more commonly 
scalable video coding), and allows to adapt the video quality to the environment. In our context, we can envision that transmission can happen with high quality if a large number of helpers (and thus a large number of bits) can be accessed by a mobile terminal. Another related technique is called `multiple description' coding \cite{goyal2001multiple}, where a file is encoded in different ways, such that each of the encoded versions can be used for a coarse reconstruction of the video, but if multiple versions are received, a higher-quality reconstruction becomes possible. To keep notation simple we will, however, ignore this aspect in the remainder of this paper, and consider video files encoded at a single quality level.

We now study a different aspect of the problem that plays a key role in our architecture:
popularity statistics of different video files. We use as an example YouTube videos, which are typically short clips, but our modeling tools are more general. It is well understood that the popularity of YouTube videos is highly unequal: a few ``viral'' videos are viewed by a very large number of people, while most others generate limited traffic. The popularity of video files is well modeled by a Zipf distribution which has 
   two main parameters: the decay constant $\gamma$, and the number of files $m$. The $\gamma$ parameter determines the ``peakiness'' of the distribution: a large $\gamma$ indicates that a very small number of files accounts for the majority of video traffic (which can be exploited well by our proposed architecture), while a small $\gamma$ indicates a more uniform distribution of popularity. The parameter $m$ describes the number of files that any of the users under consideration might want to access Ð it does not mean the number of video files available on the internet. Thus, $m$ is a function of the number of considered users $N$. An approximate model for this functional dependence is $m=\log N$; in other words, as the number of users increases, many of them will be interested in the same files as other users, but a few will want to access a different set of files. We also note that the probability of requesting a particular popular file tends to zero as $m \rightarrow \infty$ for $\gamma <1$, while it tends to a constant value if $\gamma >1$. 

\begin{figure}[t]
  \centerline{\includegraphics[width=9.7cm]{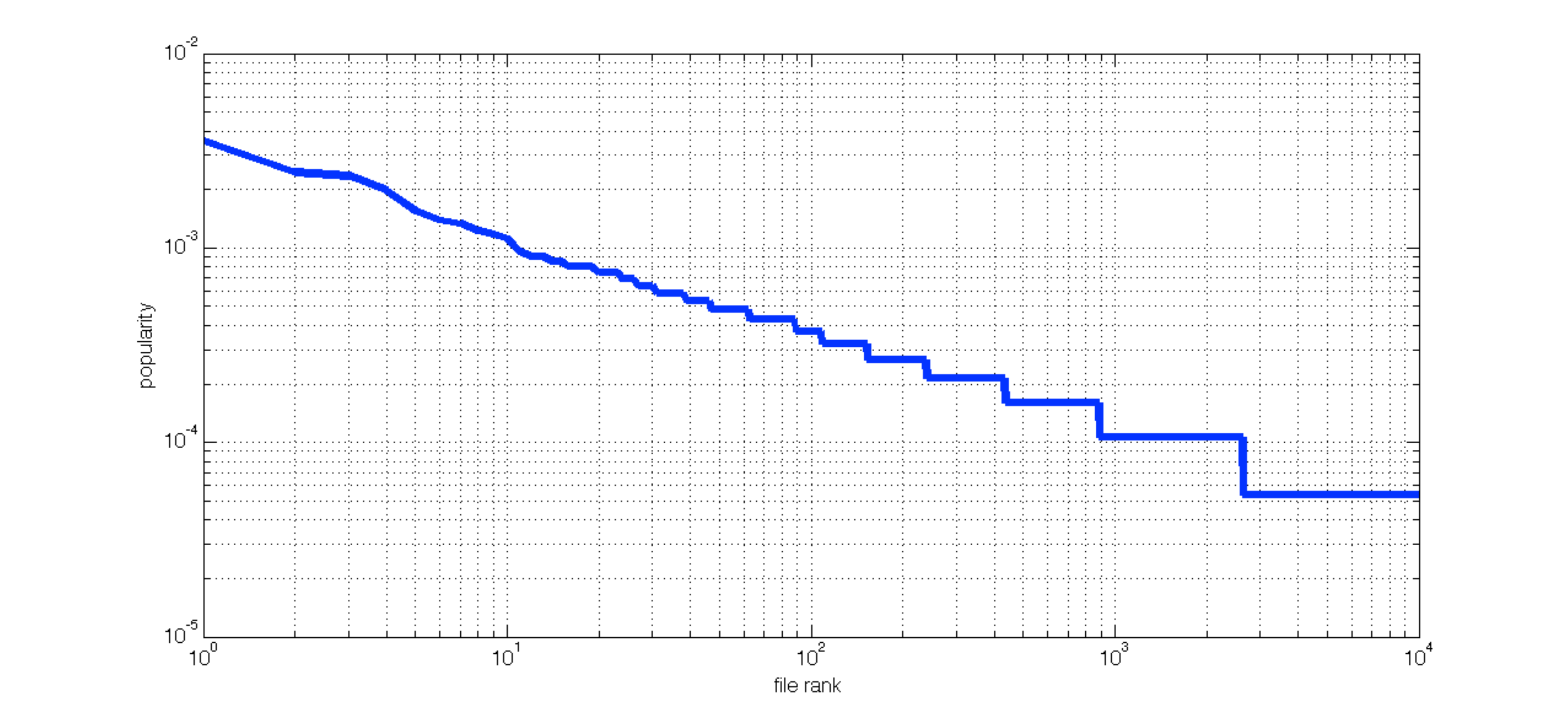}}
  \caption{An example distribution of file popularity from a wired network, namely the campus network of the University of Massachusetts at Amherst}
  \label{zipf_trace}
  \end{figure}

Figure \ref{zipf_trace}  shows an example distribution of file popularity from a wired network, namely the campus network of the University of Massachusetts at Amherst \cite{tracedata}. 
One important property that allows us to learn and exploit popularity distributions is that they change at a rather slow timescale. Typical examples include popular news, containing short videos, which are updated every $2-3$ hours, new movies, which are posted every week, new music videos, which are posted (or change popularity) about every month. Invoking a time-scale decomposition, this has two important consequences: (i) the popularity distribution of the files is approximately locally constant in time; furthermore it can be learned by the system, and thus be assumed known for our further considerations. The problem of tracking the popularity distribution and updating the content is a standard cache updating problem, for which an enormous literature exists. (ii) once the optimal content placement is determined, the operation of actually populating the helpers' caches can take place using weak backhaul links, since the cost of refreshing the helpersÕ content can be safely neglected.
\section{Femtocaching in helpers nodes}
As discussed, one general principle of our proposed architecture is to reduce the backhaul link traffic by replacing the femto base stations by small base stations that have a low-bandwidth backhaul link but high storage capacity. Here we assume that the caching helpers are placed in fixed positions in the cell and have (i) large storage capacity, (ii) localized, high-bandwidth communication capabilities which enable high frequency reuse, and (iii) low-rate backhaul links which can be wired or wireless. The key point is that if there is enough content reuse, \textit{i.e.} many users are requesting the same video content, caching can replace backhaul communication. This occurs because the most popular files are stored in the cache, and are thus always available locally to the UTs that are requesting it. Our approach is thus fundamentally different from a heterogeneous network using femto base stations, which do not have caches, and thus need to obtain any file through their backhaul network when it has been requested locally. 

In this section we consider helpers that operate in conjunction with a traditional, macrocellular base station. We consider a single cell, equipped with a macro base station (BS), serving a large number of UTs with the help of dedicated helpers. If a UT requests a file that is cached in local helpers, the helpers handle the request; the macro BS manages the requests that cannot be handled locally. Clearly, the smaller the percentage of file requests that has to be fulfilled by the macrocell, the larger the number of UTs that can be served. The central question is, of course, how much gain we can expect in real systems. Here we show that under realistic assumptions, the number of users that can be served is increased by as much as $400 -500\%$.

We start out with the case where complete files are stored in the helper stations, \textit{i.e.}, no Fountain coding is used. If the distance between helpers is large, and each UT can connect only to a single helper, it is obvious that each helper should cache the most popular files, in sequence of popularity, until its cache is full. If the helper deployment is dense enough, UTs will be able to communicate with several such helpers and each sees a distributed cache that is the union of the helpersÕ caches. In this situation, the question on how to best assign files to different helpers becomes a much more complicated issue, as shown in the example of Figure \ref{bipartite}. There are two helpers and four mobile terminals. The dashed circles centered around helpers indicate the transmission radius of helpers, \textit{i.e.}, users within a dashed circle can communicate locally with the corresponding helper. Assuming that each helper can cache $M$ files, users $U_1$ and $U_2$ would prefer helper $H_1$ to cache the $M$ most popular files since this minimizes their expected downloading time. Similarly, user $U_4$ would prefer that helper $H_2$ also caches the $M$ most popular files. However $U_3$ would prefer $H_1$ to cache the $M$ most popular files and $H_2$ the second $M$ most popular (or the opposite). This effectively creates a distributed cache of size $2M$ for user $U_3$. As can be seen, in the distributed caching problem, the individual objectives of different users may be in conflict, and we need sophisticated algorithms to find an optimum assignment.  
  \begin{figure}[htb]
  \centerline{\includegraphics[width=6.5cm]{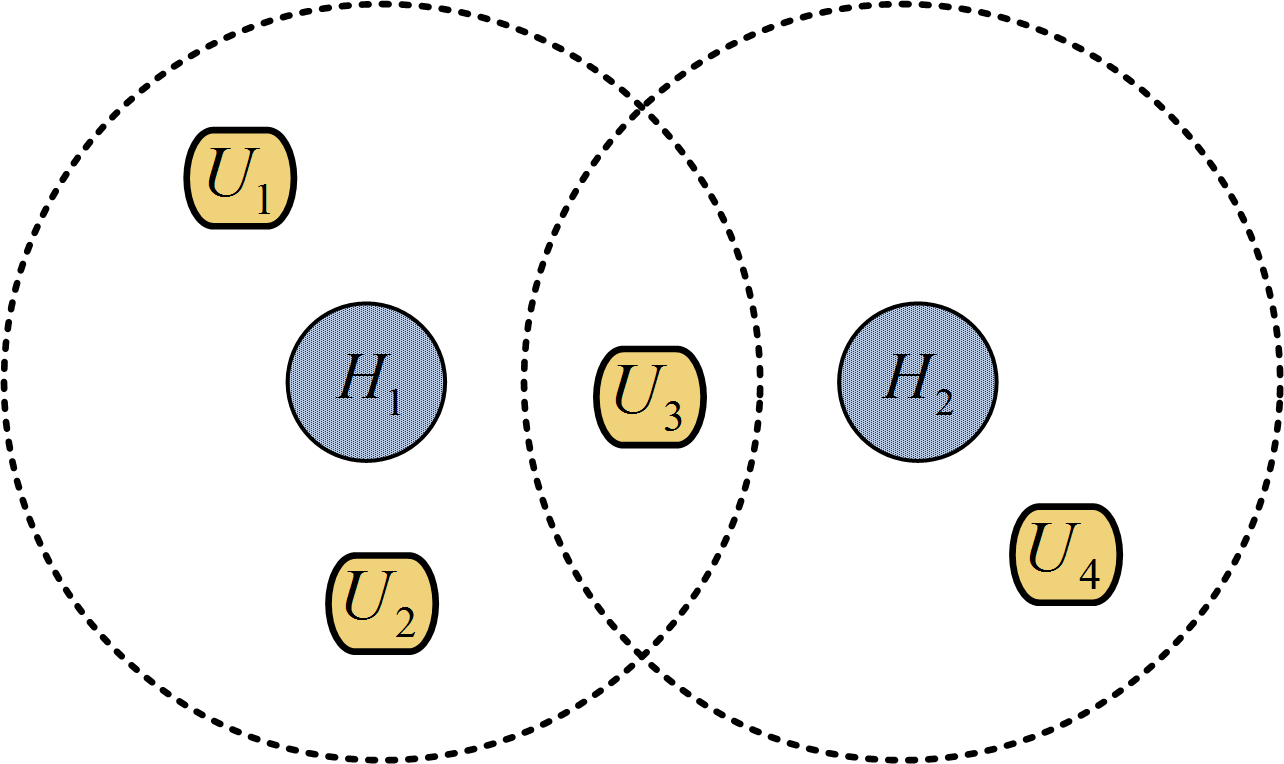}}
  \caption{Distributed Caching example: two helpers $H_1, H_2$ and four users with conflicting interests.}
  \label{fg_DistCaching}
  \end{figure}

 \begin{figure}[htb]
  \centerline{\includegraphics[width=7.5cm]{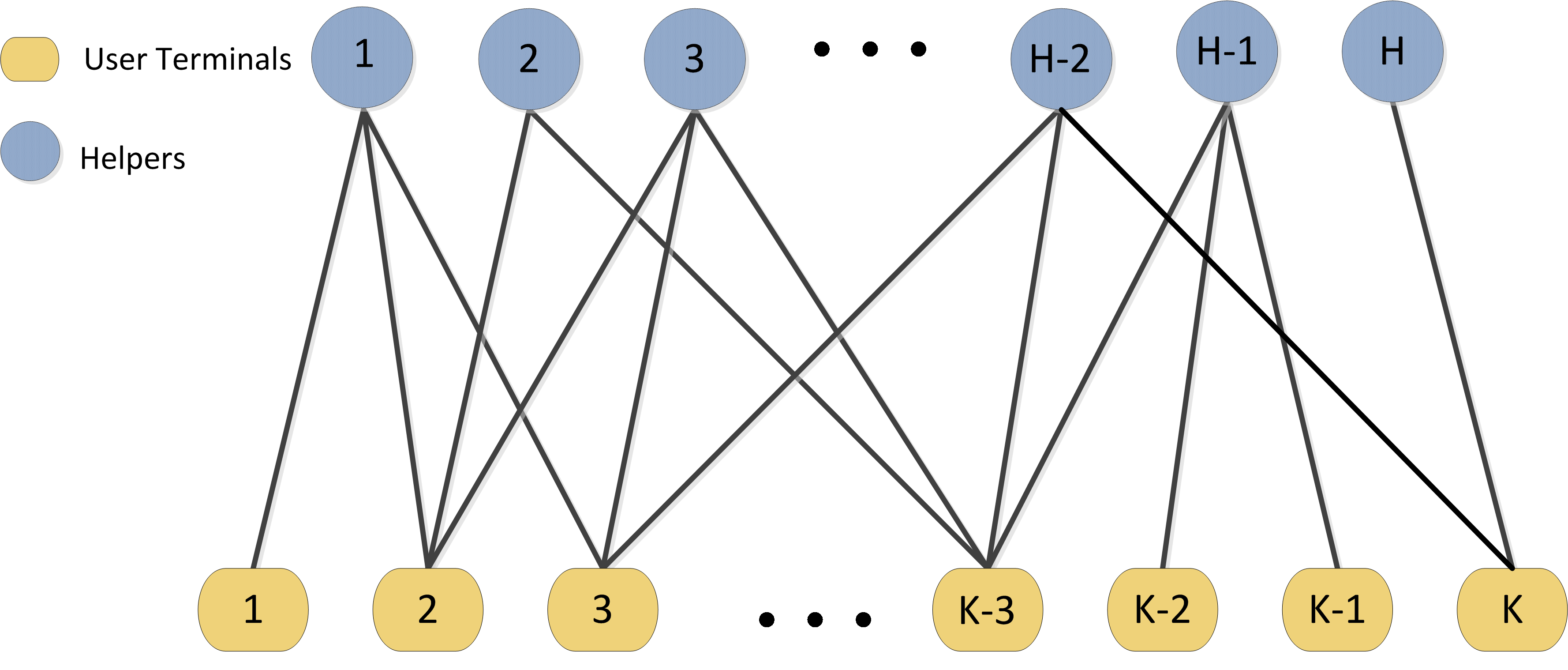}}
  \caption{Example of a connectivity bipartite graph indicating how UTs are connected to helpers.}
  \label{bipartite}
  \end{figure}

In \cite{website}, we showed how the file-assignment problem can be written as the maximization of a monotone submodular function with matroid constraints. A submodular function is, roughly speaking, a function with diminishing returns: in other words, if a file is already stored in a lot of helper nodes, the less benefit the system will gather from adding this file to yet another node. Matroids are essentially structures that generalize the concept of independence from vector spaces (where linear independence is a well-known concept) to general sets. While the exact solution is NP-hard, a result from the mathematical literature states that a simple greedy algorithm will achieve at least 50\% of the optimum value. Such a greedy algorithm fills up the set of distributed helper caches, by adding in each iteration step the file that provides the marginal best benefit for the overall cost function, namely the acceleration of file transmission. There are other algorithms that provide even better performance, but they are quite cumbersome, so we will base our simulation examples below on greedy algorithms.

Consider now the case where Fountain coding is used to simplify the combinatorial distributed caching problem. Intuitively, in the
uncoded cache, a helper could only choose to cache or not cache a file. Through coding, however, a helper can cache any
amount of coded symbols relative to a given file. In this case, the file assignment problem becomes a convex optimization problem, and through the introduction of new variables can be turned into
a linear program \cite{website}. Even though the number of involved variables is quite large, some tricks and approximations (like
bundling of video files with similar properties into groups that form the basis of optimization) can alleviate these problems.
Furthermore, the optimization can be solved in a distributed way \cite{zhang2010adaptive}. Thus, coding allows the complicated combinatorial
problem (submodular function with matroid constraints) to become a tractable convex optimization that can be solved in
reasonable time.

We now present some numerical simulation results for the gains that can be achieved by helper systems over the simple baseline system that transmits all video files via a base station. The macro-base station uses a transmission scheme similar to the fourth-generation LTE (Long-Term Evolution) standard, based on an OFDM-TDMA physical layer. The cell radius is chosen to be 400 m (which is typical in urban environments), and the users are distributed uniformly in the cell. The BS uses  Òproportional fairness schedulingÓ (PFS), where users are scheduled over time-frequency resource allocation blocks that provide good transmission quality. The helpers are operating independently, with WiFi-like links.  We assume that the data rate of the users decreases with increasing distance, but that the rate is independent of the number of users supplied by a helper station. This is a reasonable assumption if the user terminals have single antennas, and helper station has multiple antennas (\textit{e.g.}, access points following the IEEE 802.11n standard), and the number of users per helper is small. 

We target a scenario where the BS is swamped with YouTube (short videos) like video requests (following the request pattern from the above-mentioned U. Mass. Study) and consider the average downloading time
 as the metric of interest. We assume that all requests are of size 30 MB, which is reasonable assuming a screen size $640 \times 360$, flash encoding and a few minutes ( $3$ min) playback time. The simulations, analysis and conclusions also hold (in a scaled form) for larger file sizes. We define a user to be satisfied if the downloading time is below a specific threshold. 

\begin{figure}[htb]	
\centerline{\includegraphics[width=9.7cm]{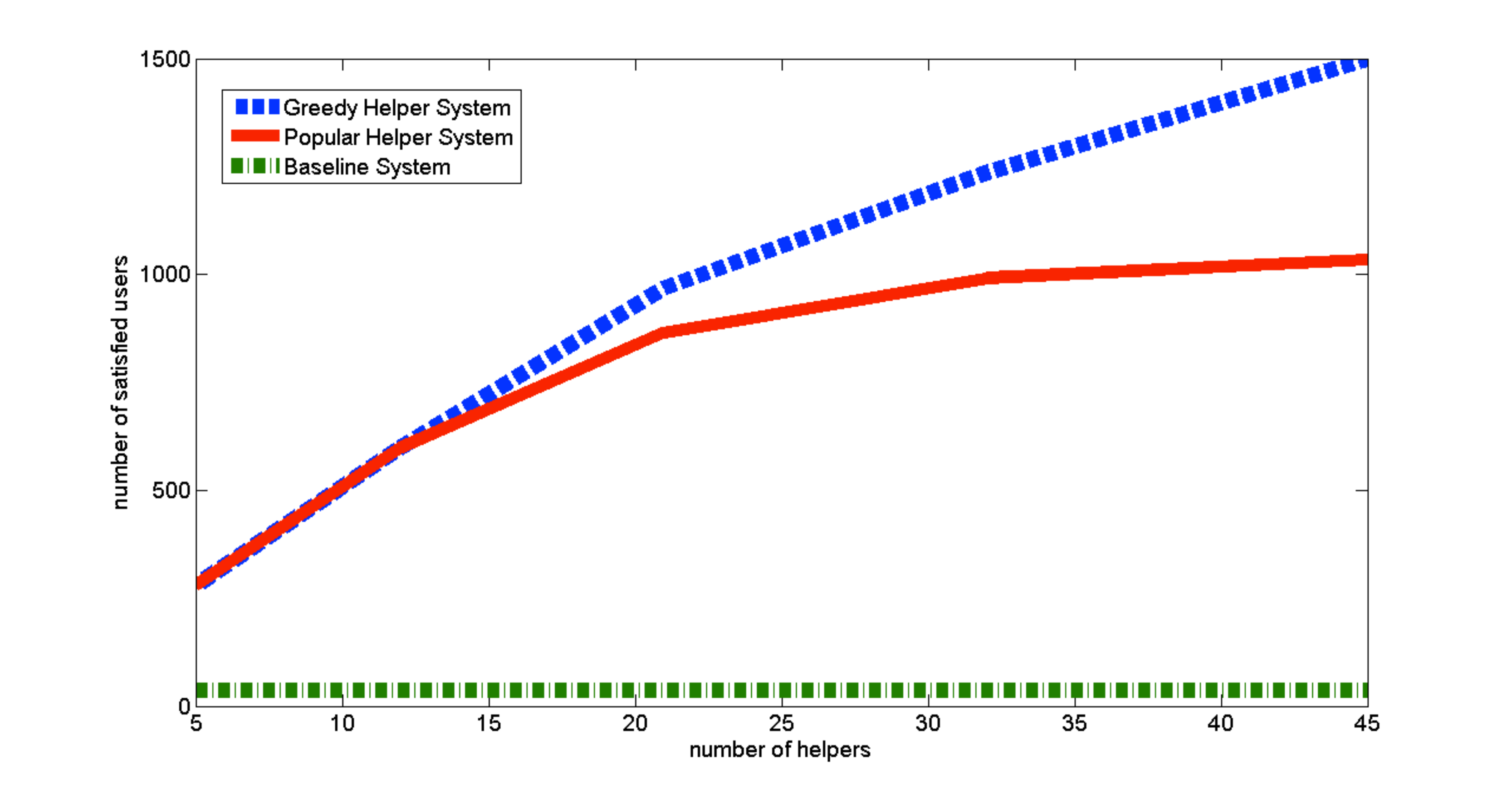}}
\caption{The number of satisfied users  versus  the number of helpers, the cache capacity of each helper
is $60GB$, QOS is $200$ seconds. }
\label{user_helper_trac2}
\end{figure}

We start by considering a situation without Fountain coding, \textit{i.e.}, complete files are stored. Figure \ref{user_helper_trac2} shows the number of satisfied users versus the number of helper nodes. We see that even for a small number of helpers ($\sim 10$), the throughput can be increased by some $\sim 500\%$. We also see that for a small number of helpers a very simple file allocation strategy (every helper stores the most popular files that fit into its cache) is as good as the greedy algorithm that (approximately) optimizes the file assignment. The reason is that in this case, each user has only connection to one helper, so that the store the most popular strategy becomes optimum. As the number of helpers increases, the performance curves start to diverge. Furthermore, the gains from adding more and more helpers start to saturate. Further simulations also demonstrate that the gains from increasing the storage capacity of each helper saturate. 
We now turn our attention to the coded system. Figure 5 compares the number of satisfied users versus the capacity of the storage capacity at the helper stations. For comparison we also show the corresponding results for the uncoded case. {We find that the performance of the coded and uncoded cases are very close to each other and we observe significant gain in both coded and uncoded systems } 

\begin{figure}[htb]	
\centerline{\includegraphics[width=9.7 cm]{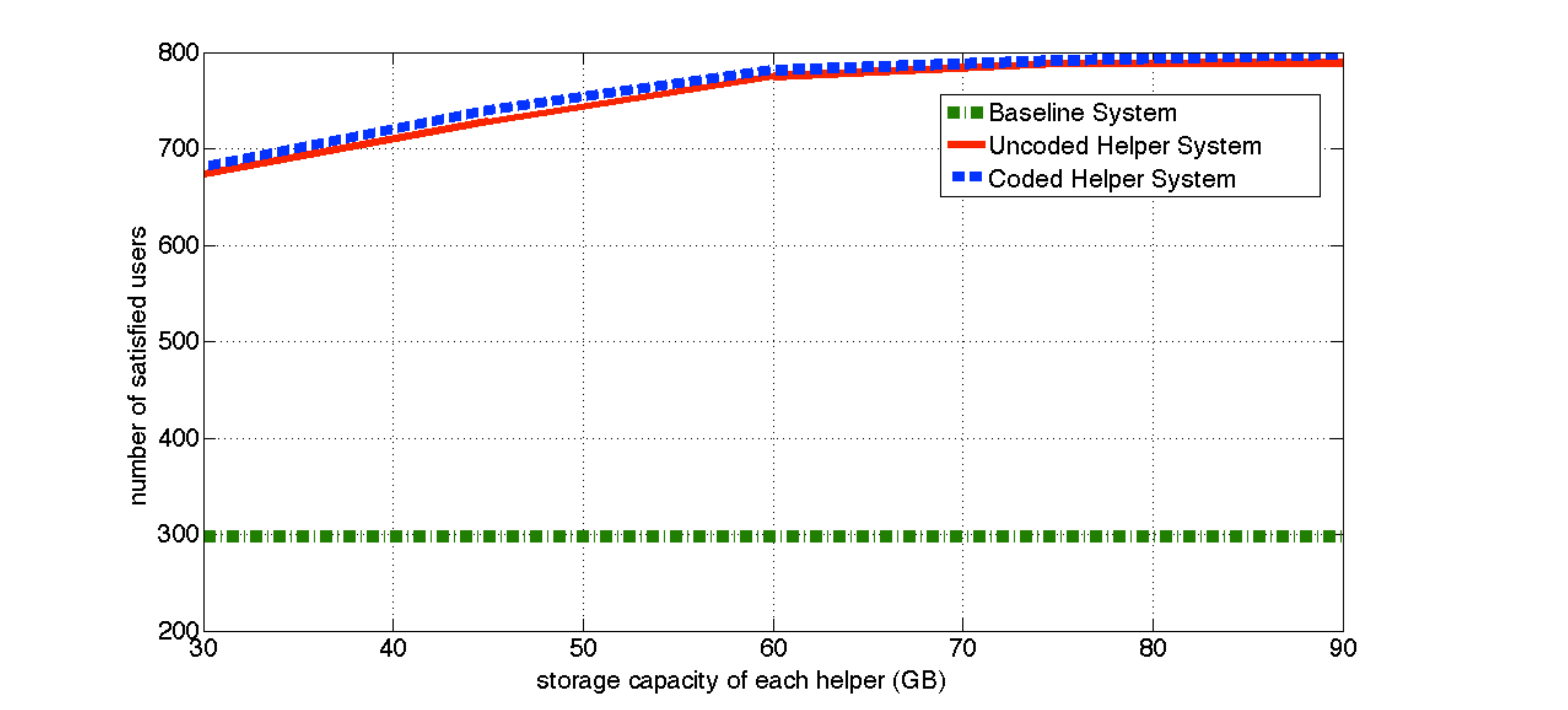}}
\caption{The number of satisfied users  versus  the cache capacity of each helper, number of helpers is 
 $32$, QOS is $200$ seconds. }
\label{comapre}
\end{figure}
To conclude this section, we discuss the various possibilities of filling the helper disks. The most straightforward method is through a wired backhaul; as discussed above, the capacity of the backhaul link does not need to be large, since the popularity of the files changes only slowly, and thus updating does not occur frequently. It is even possible to completely eschew the use of wired backhaul and have the helper stations filled up by broadcast from the macro BS during those times that its full capacity is not needed for actual user requests (\textit{e.g.}, nighttime). Yet another possibility lies in ÒoverhearingÓ file transmissions from the macro-BS to user terminals in the cell. In that case, a file has to be transmitted from the BS to a user once (when it is requested for the first time), possibly during peak hours. However, later transmissions can be done by the helper nodes. 

\section{Cellphones as caches}
Since smartphones and tablets, as well as laptops, have large hard disks built in, these devices can effectively act as mobile helper stations. They do not require special infrastructure, and have the further advantage that the number of ÒhelpersÓ is inherently concentrated in those areas where the largest demand occurs. The data transfer between ÒhelperÓ and users becomes a Òdevice-to-deviceÓ (D2D) communication. Of course, users need to be incentivized to have their stations provide help to other users: in other words, there has to be a compelling answer to the question: Òwhy should I spend {\em my} battery to provide {\em you} with faster video download? Since network operators benefit from offloading of video traffic to D2D, such incentives can be provided by network operators. They could take on, \textit{e.g.}, the form of discounts or increased data caps for participating users. Incentives could also be based on the principle of reciprocity, \textit{i.e.}, users get tokens for acting as helpers, and can then use these tokens when they want to obtain files from other mobile terminals.
 These considerations 
should be cast into game-theoretic setting which is out of the scope of this paper \cite{zhang2011peer}.

Within one cell, there are multiple pairs of possible ÒhelpersÓ and ÒusersÓ (we use this notation henceforth to describe mobile terminals that make files available and terminals that request a file; note that the role of a terminal might change with time). However, not all of those helper-user pairs can actually communicate (the connection between them might be too weak), and even among the possible communication pairs, not all can be active at the same time, since they might mutually interfere. The key question now is Òwhich helper-user pairs should be scheduled, and with what transmission parameters, to maximize D2D throughput in the cellÓ? \footnote{ instead of maximization of D2D throughput, alternative maximization of Òquality of serviceÓ (QoS) criteria can be used.}  

We assume in the following that the D2D communication is controlled by the BS, so that there is a central control unit that has knowledge about which station has what files in its cache, and also knows the channel state information (CSI) between the users. This allows a more efficient scheduling of the communications and (for the case that the D2D communication occurs in the same frequency band as the BS-to-device communication) ensures that there is no interference between the two types of traffic Ð this is essential for network operators. Furthermore, if the user terminals are stationary (nomadic, \textit{e.g.}, a laptop or tablet that is in the same location for many hours), the BS can order it to cache specific content. As we will see later on, such deterministic, centrally planned, caching can improve the efficiency of the collaboration scheme. Finally, for fast-moving terminals, the BS can use localization techniques to predict the route of the terminals, and thus pre-plan which helper-user pairs will occur in the future. 

The transmit power of the D2D communication has a critical impact on the overall system capacity. The smaller the transmit power, the smaller the region in which this communication creates interference Ð we could say that the cell size of the link is small, and we can thus put many ÒcellsÓ into one macrocell. On the other hand, a small transmit power might not be sufficient to reach a helper that has the desired file. The optimum way to assign powers is to jointly select user-helper pairs and the transmit powers optimum for these specific links. However, this optimization problem is highly complex. We here restrict ourselves to the case where every user terminal employs the same transmit power, though this one power value is optimized in the following.

To make the problem tractable, we consider the following simplified model:  a single square cell with one macro-BS serves $N$ users that are uniformly distributed in the cell. The cell is divided into smaller areas called clusters that are square with equal size. In each cluster, at most one D2D transmission is allowed, but there is no interference between two links in adjacent clusters. The cell side is normalized to $1$ and cluster side is equal to $r$, the collaboration distance. The collaboration distance is directly proportional to the transmit power (the actual dependency is determined by the pathloss). Fading effects are neglected. 

If a user requests one of the files stored in neighborsÕ caches in the cluster, neighbors will handle the request locally through D2D communication; otherwise, the BS should serve the request. As a result, the probability that D2D communication is done in clusters depends on what users store. We can imagine that users in each cluster have access to a central virtual cache (CVC) filled up with the stored files in users caches in the cluster. If the UTs are static, and the BS can control which files are stored in each UT, the best strategy is that the CVC of each cluster stores, without repetition, the $kM$ most popular files, where $k$ is the number of UTs in a cluster, and $M$ the number of files that can be cached in one terminal. 
  Note that no D2D communication is necessary if a UT finds the requested file in its own cache Ð a case we call Òself-referenceÓ.

As user terminals proliferate, and the percentage of smartphones and tablets participating in the helper network increases, the number of users per cell $n$ increases. A question is now how the number of active clusters, and with it the D2D throughput in the cell, scales with $n$.  We can then find a closed-form equation for the number of active clusters as a function of the clustersize; this serves to optimize the system parameters. Our analysis in \cite{website} shows that the result depends on the parameter of the Zipf distribution, in particular the cases $\gamma>1$ and $\gamma<1$ lead to different results (the further special case $\gamma=1$ will be omitted here for simplicity). For $\gamma>1$, the D2D throughput increases linearly with the number of users in the cell Ð in other words, the number of devices within one cluster stays constant, and the fact that a larger number of users might want to access a more diverse set of files does not impact the way the system scales. The situation is different for $\gamma<1$: in this case, the scaling occurs sublinearly, depending on the particular value of $\gamma$. 
    \begin{figure}[htb]
  \centerline{\includegraphics[width=10.2cm]{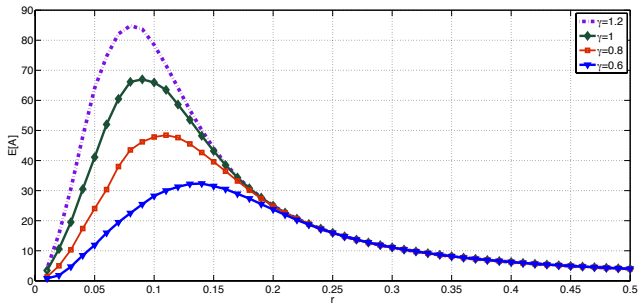}}
  \caption{The average number of active clusters versus $r$ for $n=500$, $m=1000$ and different values for $\gamma$.}
  \label{average active cluster}
  \end{figure}
  
The central control of the caching by the BS allows the very efficient file-assignment to the mobile stations mentioned above. However, if such control is not desired, and/or the mobile stations are moving around quickly, then the caching has to be done randomly. In other words, each UT will cache files according to a probability density function (pdf). This caching pdf can be chosen as an operating parameter of the network; it is noteworthy that the optimum caching pdf is not identical to the popularity pdf: imagine a situation with a very concentrated request pdf, \textit{i.e.}, a large $\gamma$. If the caching pdf were identical to the request pdf, the probability would be high that more than one node in a cluster store the most popular file -- a waste of caching space. The actually best pdf thus needs to be derived separately. In order to avoid the difficulties of optimizing a function, in one of our recent papers \cite{website} we assumed that the functional form of the caching pdf is also a Zipf distribution, but with a parameter $\gamma_1$ that can be different from the parameter $\gamma$. Still, a key insight from asymptotic analysis is that also in this case, the number of active clusters scales linearly with the number of users in the cell if both $\gamma$ and $\gamma_1$ are larger than unity. 
    \begin{figure}[t]
  \centerline{\includegraphics[width=9.7cm]{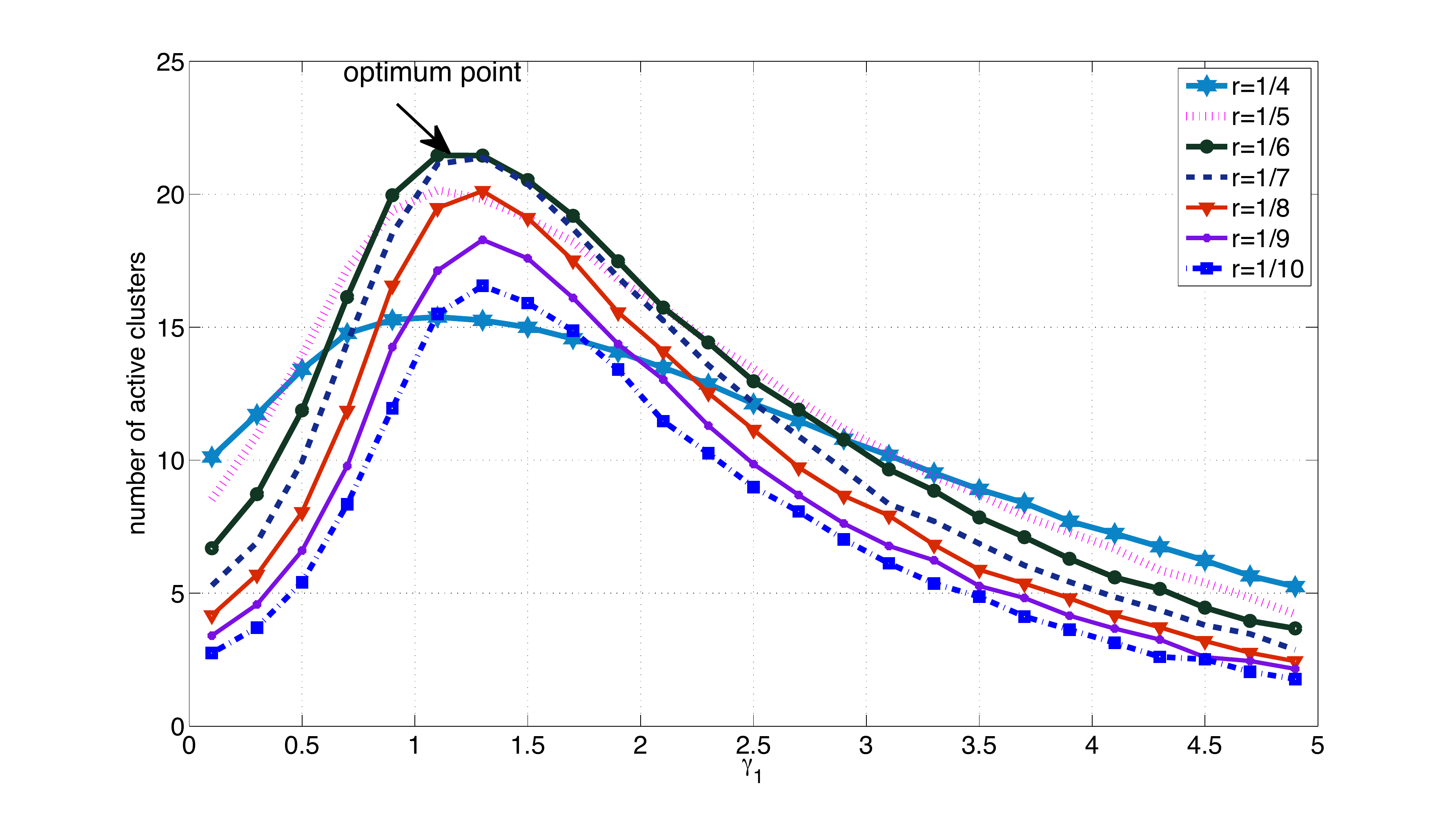}}
  \caption{The average number of active clusters versus $\gamma_!$ for random caching strategy with$n=500$, $m=1000$, $\gamma=0.6$ and different values for $r$.}
  \label{average active cluster}
  \end{figure}
  
In one of our papers, we model the wireless networks as random geometric graphs (RGG) \cite{website}. A random geometric graph $G(n,r)$ is formed by placing $n$ nodes uniformly and independently in a unit square. Two nodes can communicate with each other  through device to device (D2D) communication if the Euclidean distance between them is at most $r)$. The maximum allowable distance for D2D communication $r$ will determine the power level for each transmission.  

Above we have assumed that the UTs are already loaded with files according to a particular distribution. The question then naturally arises how the UTs can acquire the files, such that the appropriate probability distributions are fulfilled. In order to make the file distribution process efficient, UTs should cache files whose transmission (either from the BS or another helper to some requesting UT) it can overhear. But which overheard files should be cached? If the base station has control over the file storage of all users in the cell, it can order particular UTs to store particular files. In the case of static users, this enables the Òoptimum cachingÓ discussed above; in the case of users that are mobile in the cell, the BS can only ensure the appropriate caching pdf. If the BS does not control the process, then each UT should probabilistically cache or ignore files that it can overhear. We recently present an algorithm that achieves a lower bound on the average time until files are cached according to the desired distribution \cite{website}. 

\section{Outlook}
Summarizing, this article described a new architecture for increasing the throughput of cellular video transmission. By exploiting the fact that a few popular videos account for a considerable percentage of video traffic, we argue that storing those files Òclose to the usersÓ allows an unloading of a lot of traffic from the main cellular network, and thus an overall increase in throughput. The popular files can be stored either in helper nodes that have a slow (or nonexistent) backhaul and a large hard disk storage, or they can be stored on UTs, from where they can be transmitted to requesting users. 
We furthermore described the key scientific problems that need to be solved for efficient deployment of the new schemes. In the case of a helper infrastructure, the Òdistributed caching problemÓ, \textit{i.e.}, which files should be stored on which helper node, has to be solved. Depending on whether we wish to use distributed coding or keep files in a concentrated manner on the helpers, the caching problem is NP-hard (with greedy approximations coming within a constant factor of the optimum solution) or can be solved by convex optimization. 

When using the UTs as caches, many video requests can be satisfied by (base-station controlled) D2D communications, without requiring any new infrastructure. A further advantage lies in the fact that Òhelping nodesÓ automatically are concentrated in the regions where there are also the requesting devices Ð in short, other UTs. Assuming that all users have the same transmit power and thus coverage for D2D communications, we showed that there is an optimum collaboration distance that trades off the spatial frequency reuse with the probability of finding the desired file within the collaboration (communication) distance.

 These techniques look promising to achieve $1$ to $2$ orders of magnitude increase in capacity.
 Furthermore, they can be combined with a number of other improvements, such as more efficient physical-layer links, increased used of spectrum, and more efficient video coding. 

While our simulations and analytical results have shown the concept to be an extremely promising way for alleviating the capacity bottlenecks for wireless video transmission, a lot of research remains to be done. One group of topics revolves around the learning of the popularity distribution, and its possible temporal changes over time. How can popularity be assessed, and possibly predicted for the future? Accuracy of the assessment and speed of finding the distribution will have to be traded off. Furthermore, can temporal change of the popularity distribution be predicted and thus the proper loading of the caches anticipated? Could the distribution even be ÒpersonalizedÓ if the system knew user preferences as well as user mobility patterns? Somewhat related problems have been investigated in the context of video Òrecommendation systemsÓ such as the ÒNetflix challengeÓ, but the use in mobile systems offers new challenges. Also, privacy concerns need to be taken into account. 

Another group of open topics revolves around lifting some of the simplifying assumptions used in the descriptions above. For helper systems, a non-uniform distribution of helpers and/or UTs would change the optimum file assignment. For D2D communications, the use of adaptive power control might increase the reuse distance and thus the data throughput (though the scaling behavior will not change). 

A combination of the distributed caching with multi-description video coding and/or multi-level coding also poses a number of interesting challenges. Finally, relaxation of the simplified channel- and interference models, how to make all this compatible with the current technology trend of 
DASH (http over TCP, client driven adaptive streaming), and  experimental verification of the results should be done. All of those topics are currently under investigation at our laboratory, and will be reported in future work. 
\section*{Acknowledgment}
This work was supported by the Intel/Cisco VAWN (videoaware wireless networks) program. Useful discussions with Dr.
Chris Ramming and Dr. Jeff Foerster (from Intel), as well as
Prof. Mike Neely, Prof. Antonio Ortega, and Prof. Jay Kuo
(from USC) are gratefully acknowledged.
\bibliographystyle{IEEEtran}
\bibliography{ref_mag.bib}
\end{document}